\documentstyle[11pt,aaspp4]{article}

\def\degrees{\hskip-0.7mm\char'27}
\def\decdeg{\degrees\hskip-1.4mm}
\def\kkk{\mbox{kpc$^{\rm 3}$}}
\def\kmsec{\mbox{km~s$^{\rm -1}$}}

\def\BmV0{\mbox{(B-V)$^{\rm o}$}}
\def\VmK0{\mbox{(V-K)$^{\rm o}$}}
\def\MV0{\mbox{M$_{\rm V}^{\rm o}$}}

\def\etal{\mbox{{\it et al.}}}

\begin{document}
\normalsize
\raggedright
\setcounter{page}{0}

\title{A Preliminary Discussion of the Kinematics of BHB  and RR Lyrae
Stars near the North Galactic Pole.}
                 
\vspace*{0.5in}
\author{T.D.~Kinman}
\affil{Kitt Peak National Observatory\\
  National Optical Astronomy Observatories\altaffilmark{1}\\
  P.O.~Box 26732, Tucson, Arizona 85726\\ e-mail: kinman@noao.edu}
\altaffiltext{1}{ The National Optical Astronomy Observatories are 
  operated by the Association of Universities for Research in Astronomy, 
  Inc., under cooperative agreement with the National Science Foundation.}

\author{Jeffrey R.~Pier\altaffilmark{2}}
\affil{U.S.~Naval Observatory, Flagstaff Station \\ P.O. Box 1149 \\ Flagstaff, Arizona~86002-1149\\ e-mail: jrp@nofs.navy.mil}
\altaffiltext{2}{ Visiting Astronomer, Kitt Peak National Observatory, National Optical Astronomy Observatories.}

\author{Nicholas B.\ Suntzeff}
\affil{Cerro Tololo Inter-American Observatory\\
  National Optical Astronomy Observatories\altaffilmark{1}\\
  Casilla 603, La Serena, Chile\\ e-mail: nsuntzeff@noao.edu}

\author{D.L.~Harmer and F.~Valdes}
\affil{Kitt Peak National Observatory\\
  National Optical Astronomy Observatories\altaffilmark{1}\\
  P.O.~Box 26732, Tucson, Arizona 85726\\ e-mail: dharmer@noao.edu; fvaldes@noao.edu}

\author{Robert B.~Hanson, A. R.~Klemola and Robert P.~Kraft\altaffilmark{2}}
\affil{University of California Observatories/Lick Observatory, Board of Studies in 
Astronomy and Astrophysics, University of California Santa Cruz, California 95064\\e-mail: hanson@ucolick.org; klemola@ucolick.org; kraft@ucolick.org}
\
\vspace*{0.75in}
\begin{center}
Running head:  Kinematics of BHB and RR Lyrae Stars\\ 
\vspace{0.2in}
Send Proofs to:  T. D. Kinman   \\
\end{center}

\begin{abstract}

 The radial velocity dispersion of 67 RR Lyrae and BHB stars that are
more than 4 kpc above the galactic plane at the North Galactic Pole is
$\approx$110 \kmsec and shows no trend with Z (the height above the
galactic plane). Nine stars with Z $\leq$ 4 kpc show a smaller
velocity dispersion (40$\pm$9~\kmsec~) as is to be expected if they
mostly belong to a population with a flatter distribution. Both RR
Lyrae stars and BHB stars show evidence of stream motion; the most
significant is in fields RR2 and RR3 where 24 stars in the range
4.0$\leq$Z$\leq$11.0 kpc have a mean radial velocity of
$-$59$\pm$16~\kmsec~. Three halo stars in field RR 2 appear to be part
of a moving group with a common radial velocity of $-$90~\kmsec~.  The
streaming phenomenon therefore occurs over a range of spatial scales.
The BHB and RR Lyrae stars in our sample both have a similar range of
metallicity ($-$1.2$\leq$ [Fe/H]$\leq$$-$2.2).  Proper Motions of BHB
stars in fields SA~57 (NGP) and the Anticenter field (RR 7) (both of
which lie close to the meridional plane of the Galaxy) show that the
stars that have Z $<$ 4 kpc as well as those with Z $>$ 4 kpc have a
Galactic V motion that is $<$ $-$200 km/s and which is characteristic
of the halo.  Thus the stars that have a flatter distribution are
really halo stars and not members of the metal-weak thick-disk.

\end{abstract}


\section{Introduction}

The velocity dispersion of the RR Lyrae stars in the solar
neighborhood is not isotropic; an interpretation of this inequality is
that the galactic system of RR Lyrae orbits is flattened (Woolley,
1978; Hartwick, 1983; Layden 1995). This is to be expected if most of
these local halo \footnote {Following KSK, we have used the word
``halo" in a population sense to describe those field stars that are
physically similar to the stars in halo globular clusters. This seems
appropriate for the BHB and metal-weak RR Lyrae stars which are known
to occur in these clusters but not in the disk (or bulge) globular
clusters.}  stars have a flattened galactic distribution; direct
evidence for such a flattened distribution among the nearby blue
horizontal branch (BHB) stars was recently shown by Kinman, Suntzeff
and Kraft (KSK)(1994).  The more distant halo stars, like the halo
globular clusters, would be expected to have a more isotropic velocity
distribution in accordance with their more spherical spatial
distribution. Ratnatunga and Freeman (1989), however, in their study
of K-giants at the South Galactic Pole (SGP), found that the velocity
anisotropy persisted out to 25 kpc from the galactic plane and that
the velocity dispersion of these distant halo stars was only
$\approx$75 \kmsec . White (1989) argued that this result could not be
reconciled with a spherical halo and that the kinematics of the
regions sampled by Ratnatunga and Freeman might be atypical of the
whole halo. Sommer-Larsen and Christensen (1989), however, found a
significantly higher velocity dispersion ($\approx$100 to $\approx$110
\kmsec~) in their study of distant BHB stars at the galactic
poles. Some further comments on previous observations at the SGP are
given in the Appendix.

Currently there is uncertainty not only about the velocity dispersion
of the distant halo stars in the direction of the galactic poles but
also about the homogeneity of the distant halo.  Some of the
uncertainty may come from the rather small samples of stars that have
hitherto been available to derive velocity dispersions.  To help
resolve these problems we need pure samples of halo stars that are
unadulterated with thick-disk stars.  Halo tracers that meet this
criterion are the metal-poor RR Lyrae stars ([Fe/H]$\leq$$-$0.9) and
the BHB stars that have been selected by the criteria discussed by
KSK. We discuss here stars of both these types in the three Lick
Astrograph RR Lyrae fields RR~2, RR~3 and RR~4 near the North Galactic
Pole (NGP) (Kinman, Wirtanen and Janes, 1966). The metallicities
[Fe/H] of the RR Lyrae stars in these fields were determined by
Butler, Kinman and Kraft (1979). The BHB stars in RR~4 ( SA~57) are
described by KSK.  The BHB stars in the other two fields were selected
from the AF stars given by Sanduleak (1988); details will be given in
a future paper on the completion of the photometric survey in these
fields by one of us (T.D.K.).  Radial velocities have now been
obtained for the majority of the BHB and RR Lyraes (with B brighter
than $\approx$ 16.5) in these NGP fields; this paper gives a
preliminary discussion of this ongoing work.

\section{Observations and Reductions}

\subsection{RR Lyrae Stars}

The radial velocities of 38 RR Lyrae stars in these NGP fields are
taken from a larger radial velocity study of RR Lyrae stars by Pier
(1996).  The observations were obtained with the Intensified Image
Dissector Scanner (IIDS) on the Kitt Peak 2.1-m and 4-m telescopes
during the period 1983 February to 1985 May. Spectral coverage was
from 3780 to 4460 \AA~ using the 831 l/mm grating (second order) which
gave a resolution $\approx$2.2~\AA.  Radial velocities were determined
by cross-correlating spectra of the program stars against those taken
as radial-velocity templates (Tonry \& Davis, 1979; Pier,
1983). Details concerning data reduction techniques and radial
velocity measurements will be found in Pier (1996).

Multiple observations were obtained for many of the stars. The gamma
velocities were derived by correcting for the pulsation velocity using
synthetic RR Lyrae velocity curves (Woolley \& Savage, 1971; Liu,
1991); sufficiently contemporaneous ephemerides for these stars are
given by Butler, Kinman and Kraft (1979).  A comparison of the gamma
velocities of stars with repeated measures and with published
velocities indicates that the standard error in a single velocity
measurement is $\pm$15 km/s.  ~

\subsection{BHB Stars}

Some 38 BHB stars were observed in a single five night run in February
1995 with the RC spectrograph at the 4-m Mayall telescope at Kitt
Peak; 15 in field RR~4 (SA~57) and 23 in fields RR~2 and RR~3.  The
program stars cover the magnitude range 11.8$\leq$V$\leq$16.5 and
galactic latitude range 79.\decdeg~1$\leq$b$\leq$88.\decdeg~9. The
spectral resolution was 0.4\AA~ per CCD pixel
(resolution$\sim$0.8\AA~) over the waveband \mbox{$\lambda\lambda$
3880 $-$ 4580}.  The wavelength solutions were based on helium-argon
spectra taken at each slew position.  Details about the reduction
techniques and radial velocity measurements will be given in a
forthcoming paper.

Radial velocities were determined for the program stars by standard
cross correlation techniques (Tonry \& Davis 1979) as described in
KSK.  We used field horizontal branch stars with well-determined
velocities by Green \& Morrison (1993) for the template spectra and
for the nightly zero-points of the velocity system.  The mean
difference (our radial velocities {\it minus} those of Green \&
Morrison) for 13 spectra of HD 60778, HD 74721 and HD 161817 taken
over the five night run is $-$1.0 \kmsec and the r.m.s. value of a
single difference is $\pm$1.4 \kmsec. These results and the
cross-correlation of the comparison spectra show that the nightly
velocity zero point is accurate to 1.0 \kmsec.

The velocities were measured using the technique described by KSK. Two
velocities were measured for each star: v(weak) from the correlation
of the weak lines (with the hydrogen Balmer lines masked out), and
v(H) based on the correlation of the heavily filtered Balmer lines
(with the weak lines masked out).  The r.m.s. difference between the
two velocity measurements was 2.5 \kmsec\ for 77 spectra of bright
field giants, and 4.3 \kmsec\ for 56 spectra of program stars. The
nightly mean difference between the two systems of velocity
measurements was always less than 1.5 \kmsec.

As discussed in KSK, the correlation based on the weak lines produces
a sharper correlation peak (and a somewhat more accurate velocity)
than that based on the hydrogen lines for high S/N data or stars which
have strong metallic lines.  For 90 \% of the program stars, the
weak-line correlation peak was well defined and we could measure
v(weak) .  For the rest of the stars, we used the hydrogen line
correlation peaks to measure the stellar velocity.  The velocity
errors are $\sim 4$ \kmsec for most program stars, rising to 8 \kmsec
for the fainter stars.

\section{Discussion}
\subsection{Absolute Magnitudes and Distances}

Absolute magnitudes (M$_{\rm v}$) were derived for the BHB stars using
the cubic expression in (B$-$V)$_{\rm 0}$ given by Preston {\it et
al.} (1991) which assumes M$_{\rm v}$ equals +0.6 for the RR Lyrae
stars.  The absolute magnitudes of the RR Lyrae stars were calculated
from:

\begin{equation}
 M_{\rm v} = 1.05 + 0.29[Fe/H]  
\end{equation}
  
which is the mean of the direct and inverse relations given by Feast
(1995).  A mean B$-$V of +0.37 was assumed for the RR Lyrae stars and
their distances were calculated from the mean B magnitudes given by
Kinman \etal~(1966).  Equation (1) gives M$_{\rm v}$ = +0.59 for
[Fe/H] = $-$1.6 (a characteristic metallicity for our sample). This is
consistent with our assumed M$_{\rm v}$ for the BHB stars. Other
recent estimates of the M$_{\rm v}$ of the RR Lyrae stars
(e.g. Carney, Storm \& Jones, 1992) lie within $\pm$0.20 mag of the
value given by equation (1). The corresponding 10\% uncertainty in our
distances does not affect the conclusions of this paper.
 
Following KSK, the height Z above the galactic plane was calculated
for each star assuming no reddening. The radial velocities were
corrected to the local standard of rest and for a solar galactic
rotation of 220 \kmsec.  The adoption of a larger solar galactic
rotation (e.g. 275 \kmsec, Majewski (1992)) would not materially
change our conclusions.

\subsection{Radial Velocities}

The mean radial velocities and their dispersions were calculated separately 
for the RR Lyrae stars and BHB stars and are given as a function of Z
in Table 1 and Fig.1. For Z $<$ 4 kpc, we would expect most of the
stars to come from the flatter halo and have a lower radial velocity
dispersion.  We find that the nine stars with Z $<$ 4 kpc have a
velocity dispersion of only 40$\pm$9~\kmsec~.  The BHB stars with Z
$<$ 4 kpc at the SGP also show a lower velocity dispersion (for
details see Appendix).  For Z $>$ 4 kpc, both RR Lyrae stars and BHB
stars show a radial velocity dispersion of $\approx$110 \kmsec~ in
agreement with the results of Sommer-Larsen \etal (1989) but not with
those of Ratnatunga and Freeman (1989).  The low velocity dispersion
found by Ratnatunga and Freeman may arise partly from adulteration
with disk stars (see discussion in Sec. 9 and Fig. 19 of KSK).

The {\it mean} radial velocities of both the RR Lyrae and the BHB
stars in the NGP fields are negative in the distance range 4 to 14 kpc
(Table 1) \footnote{ The SGP stars do not show this effect. As noted
in the Appendix, the 15 BHB stars listed by Flynn \etal (1995) have a
mean velocity of $-$4$\pm$22 \kmsec~ while the 19 BHB stars with Z $>$
4 kpc in Pier's sample have a mean velocity of +15$\pm$21 \kmsec~.}.
In particular, the twenty four RR Lyrae and BHB stars in fields RR 2
and RR 3 that lie in the range in Z of 4.0 to 11.0 kpc have a mean
radial velocity of $-$59$\pm$16 \kmsec.  It is very interesting that
in a 0.3 square degree field in SA~57, Majewski, Munn and Hawley
(1994, 1996) have found stars in the radial velocity range $-$48 to
$-$86 \kmsec.  Their sample contains mostly subdwarfs near the
main-sequence turn-off with 5 $\leq$Z$\leq$10 kpc and are in a
relatively small volume of space($\sim$0.03~\kkk~).  Our mean negative
velocity in SA~57 of $-$34$\pm$27 \kmsec~ for the sixteen RR Lyrae and
BHB stars in the range in Z of 4.0 to 11.0 kpc is less significant
although there are five of these stars in the Z range of 4.6 to 6.2
kpc that have a mean radial velocity of $-$108$\pm$37 \kmsec~.  These
last 5 stars have metallicities [Fe/H] in the range $-$1.4 to $-$2.1
and occupy a volume some 6 times larger than the volume of the
Majewski \etal~field.

There is a peak in the radial velocity distribution in field RR~2 at
$-$90 \kmsec. In particular, there are three stars (one RR Lyrae and
two BHB) close to 12$^{\rm h}$:09$^{\rm m}$:05$^{\rm s}$,
+33$^{\circ}$:03$^{\prime}$ (1950) that (within the measuring errors)
have the same radial velocity; a summary of their properties is given
in Table 2. These three stars have a mean height above the galactic
plane of $\sim$10.0 kpc and occupy $\sim$0.05 \kkk~ $-$ comparable to
the volume occupied by the stars in the Majewski \etal~field.  It
seems likely that these three stars constitute a ``moving group" that
had a common origin (for references to possible high-velocity
moving-groups, see Majewski \etal~(1994)); it is very desirable to
check whether they also share a common proper motion.

The crossing time of the galactic orbits of halo stars is much less
than the likely age of halo stars so it has sometimes been concluded
that the halo is reasonably well mixed (Sommer-Larsen \& Zhen, 1990;
Binney, 1994).  Indeed, this may perfectly true of the majority of
halo stars.  The detection of a few stars with a common motion in a
small volume of space may result from the fortuitous discovery of the
remnants of a cluster in the process of dissolution.  The volume
included in fields SA~57, RR 2 and RR 3, however, is of the order of
some tens of a cubic kiloparsec. The existence of streaming motions in
a volume this size suggests a larger scale phenomenon in which the
stars have only entered the galaxy in relatively recent times and/or
that their orbits have never taken them into regions in the Galaxy
where organized motion would be rapidly destroyed.

The picture given by the RR Lyrae and BHB stars will not be the same
as that given by the subdwarfs.  The RR Lyrae and BHB stars make up a
relatively small fraction of the stellar halo; consequently, although
they are useful tracers of large scale structure, they generally do
not allow a very fine spatial resolution.  The halo subdwarfs (more
than a hundred times more numerous) can be used to sample much smaller
volumes of space and are useful for investigating the inhomogeneities
in the halo that apparently exist on smaller spatial scales.

Majewski, Munn and Hawley (1996) consider the possibility that the
streaming that they find in SA~57 may be associated with the
disintegrating Sagittarius Dwarf that has recently been found by
Ibata, Gilmore and Irwin (1994). This seems quite possible.  Among
recently suggested orbits for the Sagittarius Dwarf, (e.g. Velazquez
and White (1995) perigalacticon 10~kpc and apogalacticon 52~kpc;
Lynden-Bell and Lynden-Bell (1995) perigalacticon 12 kpc and
apogalacticon 36 kpc), our data show best agreement in radial velocity
and galactic location with orbit D of Lin, Richer, Ibata and Suntzeff
(1995).  The determination of proper motions and hence space motions
and orbits for the BHB and RR Lyrae stars in RR 2 and RR 3 is clearly
very desirable if we are to be certain that these stars are indeed
part of a stream that includes the Sagittarius Dwarf Galaxy.

\section{Proper Motions}

Preliminary absolute proper motions have been determined for 21 
BHB stars in the SA~57 field (at the NGP) and 14 BHB stars in the RR~7
Anticenter field of KSK, as part of a larger study by Hanson and
Klemola (1995), using the 51-cm Carnegie Double Astrograph at Lick
Observatory and the photographic plate collection of the Lick Northern
Proper Motion (NPM) program (Klemola, Jones, and Hanson 1987).

Four third-epoch blue (103a-O) plates were taken by Klemola in
February and April 1995.  These were measured along with first-epoch
NPM plates taken between 1950 and 1953.  Details of the NPM observing,
plate measurement, and reduction procedures are given by Klemola,
Jones, and Hanson (1987).  The proper motions were determined in an
absolute reference frame defined by 80--100 galaxies ($16 \leq B \leq
18$) per field.  Epoch spans well over 40 years allow the measurement
of proper motions with random errors $\approx 0\farcs 3\ {\rm
cent}^{-1}$, corresponding to a transverse velocity error $\approx 15\
{\rm km~sec^{-1}~kpc}^{-1}$.

The proper motions of these BHB stars were converted to transverse
velocities using the distances (kpc) given by KSK. Both these fields
lie close to the meridional plane of the Galaxy which passes through
the Sun, the Galactic Center and the North Galactic Pole. To a first
approximation, therefore, the galactic rotation vector (V) is almost
entirely given by these transverse velocities.  Fig. 2 shows this V
vector on a plot of the transverse velocities for (a) SA~57 and (b)
the RR~7 field. The filled circles show stars with Z $\leq$ 4 kpc and
the open circles stars with 4$\leq$Z$\leq$6 kpc and show that both
these groups of stars lag the galactic rotation of the Sun by at least
200 \kmsec~.  Thus the stars that have Z $\leq$ 4 kpc which must
belong predominantly to the flatter halo of KSK must indeed be
identified with the halo and not with the metal-weak ``thick-disk"
(Morrison, Flynn \& Freeman, 1990) whose rotation with respect to the
Sun should only be about 40 \kmsec~.

\section{Conclusions}

The velocity dispersion of a sample of 67 RR Lyrae and BHB stars in
three fields at the NGP and more than 4 kpc above the galactic plane
is $\approx$110 \kmsec~; this sample shows no trend of velocity
dispersion with Z. The 9 stars with Z $\leq$ 4 kpc show a smaller
velocity dispersion (40$\pm$9 \kmsec~) as is to be expected if they
belong to a spatially flatter distribution.  The mean velocity of both
the RR Lyrae and BHB stars shows evidence of stream motion in the
halo. The most significant systematic motion is that in fields RR 2
and RR 3 where 24 stars in the range 4.0$\leq$Z$\leq$11.0 kpc have a
mean radial velocity (corrected for galactic rotation and to the local
standard of rest) of $-$59$\pm$16 \kmsec~. Three halo stars in field
RR 2 occupy a volume of $\sim$0.05 \kkk~about 10 kpc above the
galactic plane and have the same radial velocity ($-$90\kmsec~). The
streaming therefore occurs over a range of spatial scales. That on the
smaller scale is comparable both in sense and size with that found by
Majewski, Munn and Hawley (1994) (1996). The possibility that these
streaming stars belonged to the newly discovered Sagittarius Dwarf
Galaxy is not in disagreement with some recent orbits that have been
calculated for this galaxy and this requires further investigation.

Preliminary proper motion data for BHB stars in SA~57 and the
Anticenter field RR~7 show that the stars with Z $\leq$ 4 kpc have a
galactic rotation similar to the BHB stars in the more distant halo;
consequently these stars of the flatter halo are really halo stars and
not members of the metal-weak thick-disk of Morrison, Flynn and
Freeman (1990).

\acknowledgements

We would like to thank Ed Carder and Jim DeVeny for their help with
the IIDS and Dave Summers and Kurt Loken for their assistance at the
4-m telescope when the BHB spectra were taken. We should also like to
thank Elizabeth Green for providing us with radial velocity
measurements of the bright BHB stars and Taft Armandroff for bringing
the paper by Velazquez and White to our attention.  The Lick Northern
Proper Motion program is supported by National Science Foundation
grant AST 92-18084.

\appendix

\begin{center}
APPENDIX
\end{center}

The 15 BHB stars near the SGP with Z$\leq$ 16 kpc given by Flynn,
Sommer-Larsen, Christensen and Hawkins (1995) have a radial velocity
dispersion ($\sigma$) relative to the mean of $\approx$82 \kmsec. If
the most distant four stars are omitted from this sample, however, the
$\sigma$ of the remaining 11 stars is $\approx$91 \kmsec.  Pier (1984)
found that the $\sigma$ of 31 BHB stars near the SGP was only
$\approx$80 \kmsec; the dispersion for the 24 of these BHB stars that
are more than 4 kpc from the plane is $\approx$87 \kmsec, while the 7
with Z $<$ 4 kpc have a dispersion of $\approx$55 \kmsec. If we
exclude the stars with B$-$V $\leq$$-$0.03 and H$\delta$-width $\geq$
30\AA~ (those least likely to be BHB stars), the dispersions are
respectively $\approx$90 \kmsec (19 stars) and $\approx$48 \kmsec.
These are quite small samples and their uncertainties are
correspondingly high; the r.m.s. error in $\sigma$ equals $\sigma$
divided by the square root of twice the number of objects, so $\sim$50
objects must be observed to get an error of $\pm$10 \kmsec~ in the
dispersion.

\clearpage

\figcaption[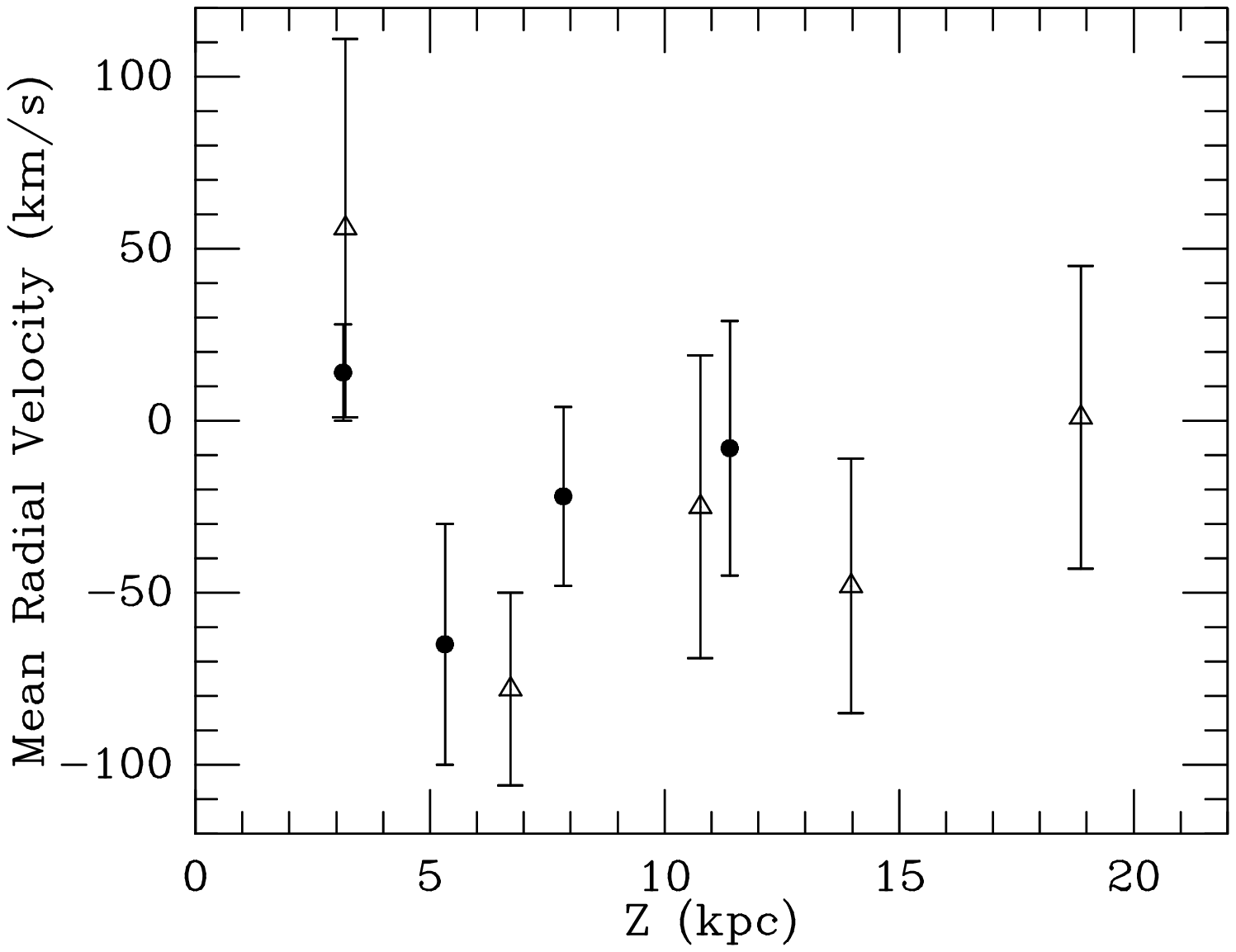]{Mean radial velocities (in \kmsec~) of program stars
as function of their distance Z (in kpc) above the Galactic plane. The
BHB stars are shown by filled circles and the RR Lyrae stars by
triangles.  \label{f1}}

\figcaption[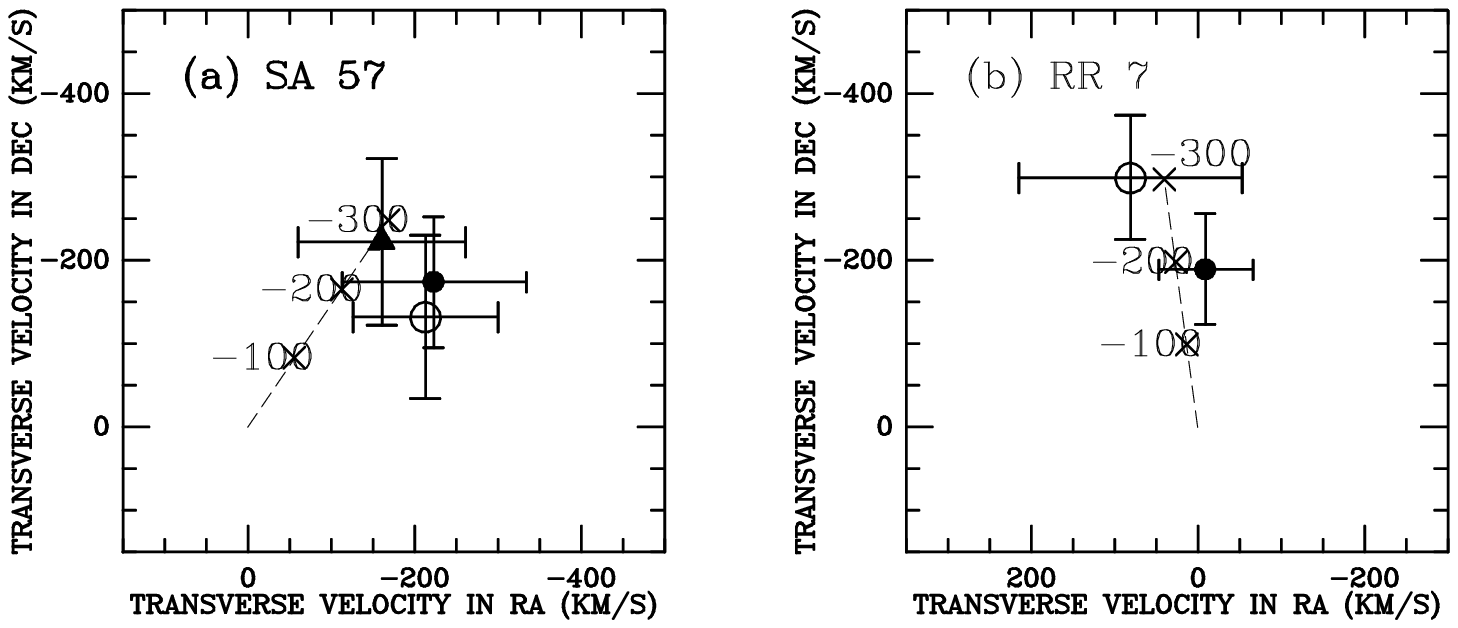]{The galactic rotation V-vector on a plot of the
transverse velocities (in \kmsec~) in declination (ordinate) and
R.A. (abscissa) for (a) SA~57 and (b) RR~7.  Stars with Z $\leq$ 4 kpc
are shown by filled circles; stars with 4 $\leq$ Z $\leq$ 6 kpc by
open circles; stars with 6 $\leq$ Z $\leq$ 12 kpc (SA~57 only) by a
filled triangle.  The dashed line shows the orientation of the V
vector in each field; the numbers show the size of the vector in
\kmsec.  The V value at any point is given by its perpendicular
projection onto the V-axis.  \label{f2}}

\clearpage

\begin{figure}
\plotone{fig1.eps}
{\center Kinman et al. Figure~\ref{f1}}
\end{figure}

\clearpage

\begin{figure}
\plotone{fig2.eps}
{\center Kinman et al.  Figure~\ref{f2}}
\end{figure}

\clearpage

\begin{table}
\dummytable\label{t1}
\end{table}

\begin{table}
\dummytable\label{t2}
\end{table}

\begin{table}
\dummytable
\end{table}

\noindent{\sc TABLE}~\ref{t1}.  Mean Radial Velocities and their
Dispersions for RR Lyrae and BHB stars in Three Fields near the North
Galactic Pole.

\noindent{\sc TABLE}~\ref{t2}. Collected Data for RR Lyrae and BHB
stars in Field RR 2 which may have a common origin.

\end{document}